\documentclass[aps,prx,twocolumn,superscriptaddress,floatfix,longbibliography]{revtex4-2}

\usepackage{amsmath}
\usepackage{amssymb}
\usepackage{graphicx}
\usepackage{tabularx}
\usepackage{makecell}
\usepackage{bm}
\usepackage{mathbbol}
\usepackage{mathrsfs}
\usepackage{xcolor}
\usepackage[colorlinks=true,linkcolor=blue,citecolor=blue,urlcolor=magenta]{hyperref}
\usepackage{placeins}
\usepackage{comment}
\usepackage{siunitx}
\usepackage{multirow}

\begin{document}

\title{Controlling photothermal forces and backaction in nano-optomechanical resonators through strain engineering}

\author{Menno H. Jansen}
\affiliation{Center for Nanophotonics, AMOLF, Amsterdam, The Netherlands}
\author{Cauê M. Kersul}
\affiliation{Centro de Tecnologia da Informação Renato Archer, Campinas, Brazil}
\author{Ewold Verhagen}
\email{verhagen@amolf.nl}
\affiliation{Center for Nanophotonics, AMOLF, Amsterdam, The Netherlands}

\date{December 23, 2025}

%

\begin{abstract}
  In micro- and nanoscale optomechanical systems, radiation pressure interactions are often complemented or impeded by photothermal forces arising from thermal strain induced by optical heating.  We show that the sign and magnitude of the photothermal force can be engineered through deterministic nanoscale structural design, by considering the overlap of temperature and modal strain profiles. We demonstrate this capability experimentally in a specific system: a nanobeam zipper cavity by changing the geometry of its supporting tethers. A single design parameter, corresponding to a nanoscale geometry change, controls the magnitude of the photothermal backaction and even its sign. These insights will allow engineering the combined photothermal and radiation pressure forces in nano-optomechanical systems, such that backaction-induced linewidth variations are deterministically minimized if needed, or maximized for applications that require cooling or amplification at specific laser detuning.
\end{abstract}

\maketitle

\section{Introduction}
Cavity optomechanics offers powerful mechanisms for controlling mechanical resonators via light, with applications ranging from precision sensing \cite{krause_high-resolution_2012, guha_force_2020, wu_nanocavity_2017, yu_cavity_2016, gavartin_hybrid_2012} to quantum information processing \cite{barzanjeh_optomechanics_2022, riedinger_non-classical_2016, zivari_-chip_2022, bozkurt_mechanical_2025}. Specifically, the dissipation and occupation of mechanical resonators can be controlled through dynamical backaction, leading to cooling or amplification \cite{kippenberg_analysis_2005, wilson-rae_theory_2007, marquardt_quantum_2007, schliesser_resolved-sideband_2008}. By harnessing this effect, ground state cooling of macroscopic mechanical resonators has been demonstrated \cite{teufel_sideband_2011, chan_laser_2011}. Traditionally, radiation pressure and, for higher frequency mechanics, electrostriction have been considered the primary forces mediating the interaction between optical and mechanical degrees of freedom \cite{aspelmeyer2014review}. However, the confinement of the optical field used to achieve large optomechanical coupling means that in many systems there are significant photothermal forces, arising from the strain produced by thermal expansion due to local heating \cite{favero_optomechanics_2009, camacho_characterization_2009, restrepo_classical_2011,woolf_optomechanical_2013,guha_high_2017,hauer_dueling_2019,shakespeare_thermal_2024,ortiz_influence_2024}. A better understanding of photothermal forces could allow their maximization or minimization, expanding the level of control in optomechanical application scenarios in both the classical and quantum domains.

In photonic cavities, backaction that modifies mechanical response arises from optical forces that depend on mechanical displacement. Photothermal forces differ fundamentally from traditional optomechanical forces in how they affect mechanical dissipation, as a result of the different temporal dynamics of the mechanisms involved. The in-phase component of these forces affects the mechanical frequency, while the out-of-phase component can affect the dissipation. The ratio between these components depends on their delay in responding to a change in mechanical position \cite{aspelmeyer2014review}. Whereas radiation pressure relies on the photon lifetime to introduce a delay, the photothermal out-of-phase component arises from slower thermal time constants. This makes it particularly potent in low-frequency mechanical systems in the unresolved sideband regime, where it can dominate the linewidth variation from optomechanical forces \cite{metzger_optical_2008}. The photothermal force has even been argued to be able to achieve ground-state cooling in absorption-limited cavities \cite{restrepo_classical_2011}. Additionally, optical heating can introduce frequency shifts with a detuning dependence that differs from that expected from pure optomechanical backaction.

Cancellation of the photothermal force has been investigated using an optically levitated mirror \cite{qin_cancellation_2022}, but the sensitive interplay of optical, thermal, and mechanical strain fields in integrated systems opens avenues for more precise engineering. Recently, it has been argued that photothermal backaction can be predicted by evaluating the spatial overlap between thermal fields and the strain profiles of mechanical eigenmodes, and shown that this corresponds well to experimentally found values \cite{primo_accurate_2021}.

The question is whether such understanding of the origin of this force can be leveraged to deterministically manipulate it, controlling its sign and magnitude at will. In this work, we show that by engineering the geometry of a nanoscale optomechanical device, the sign and magnitude of the photothermal force can be deterministically controlled. This can be achieved through subtle design changes that do not affect the essential optical and mechanical performance of the device. To this end, we first briefly discuss the theory of photothermal effects in optomechanical devices in terms of modal fields. We then use this to simulate the photothermal forces in a model system, a nanobeam zipper cavity, revealing symmetries that can be broken and exploited to manipulate the photothermal force. Finally, we show experimentally that breaking strain symmetry indeed allows the tuning of the photothermal force. We observe that optomechanical dissipation depends strongly on design, and can in fact be completely reversed. This opens a new degree of freedom for the design of nano-optomechanical systems, that could be employed in many different material and cavity design platforms.

\section{Results and discussion}

\subsection{Photothermal and optomechanical backaction}
\label{subsec:Backactions}

\begin{figure}[h!]
  \includegraphics[width = \columnwidth]{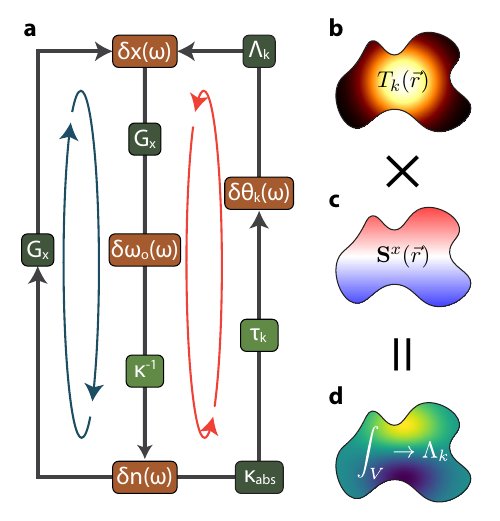}
  \caption{\small{\textbf{a} Schematic of the optomechanical (dark blue) and photothermal (red) backaction cycles. \textbf{b}, \textbf{c}, and \textbf{d} Schematic examples of a temperature profile $\tilde{T}_k(\vec{r})$, strain profile ${\mathbf{S}}^x(\vec{r})$, and the integrand of \autoref{eq:Lambdak} resulting from their multiplication.}}
  \label{fig:BackactionLoop}
\end{figure}

In an optomechanical device, there are multiple pathways for position fluctuations of the mechanical resonator to generate a force that acts back on the resonator. Two of these, through the optomechanical and photothermal force, are schematically introduced in \autoref{fig:BackactionLoop}a. Both photothermal and optomechanical backaction cycles can be understood starting with optomechanical transduction; fluctuations in the mechanical displacement, denoted in the frequency domain by $\delta x(\omega)$ in \autoref{fig:BackactionLoop}a, lead to fluctuations in the frequency of the optical mode $\delta\omega_\mathrm{o}(\omega) = G_x \delta x(\omega)$, where $G_x$ is the optomechanical coupling. Considering that the optical mode is driven by a laser at constant frequency, the resulting detuning fluctuations, $\delta \Delta(\omega)$,  will affect the number of photons, $\delta n_\mathrm{ph}(\omega) = \Psi(\omega,\Delta) G_\mathrm{x} \delta x(\omega) n_\mathrm{ph} $, where $\Psi(\omega,\Delta) = ((\Delta+ \omega) +i\kappa/2)^{-1} + ((\Delta - \omega) - i\kappa/2)^{-1}$ is the symmetrized optical susceptibility, whose imaginary part is related to the optical delay given by the inverse of the optical mode decay rate, $\kappa^{-1}$. In the case of optomechanical backaction, these photon number fluctuations generate a fluctuating force that is also proportional to $G_x$, being the converse effect to the transduction. Upon joining all steps, the overall optomechanical backaction force can be written as:

\begin{equation}
  F_\mathrm{om}(\omega) = \hbar G_\mathrm{x}^2 \Psi(\omega, \Delta) \delta x(\omega) n_\mathrm{ph}.
  \label{eq:OptomechanicalBackaction}
\end{equation}

In addition to the optomechanical force, $\delta n_\mathrm{ph}(\omega)$ leads to fluctuations in the absorbed number of photons, proportional to the optical absorption rate $\kappa_\mathrm{abs}$, which in turn causes temperature fluctuations across the device. In the treatment of the photothermal effects, one should perform a multimodal expansion of the thermal field, which is discussed more thoroughly in \ref{suppsec:Theoretical model}.
Considering here only the $k$-th thermal mode, the amplitude of the fluctuations can be written as $\delta \theta_k (\omega) = R_k^\theta \frac{\chi_k^\theta(\omega)}{\tau_k} \delta n_\mathrm{ph} (\omega)$, where $R_k^\theta$ is a thermal resistance that depends on the overlap between the optical and thermal fields, $\tau_k$ is the thermal response time and $\chi_k^\theta(\omega) = \frac{1}{1/\tau_k - i\omega}$ is the thermal susceptibility, responsible for introducing the extra thermal delay, given by $\tau_k$. Finally, $\delta \theta_k (\omega)$, leads to a contribution to the photothermal force $F_{\mathrm{pt},k}(\omega) = \Lambda_k \delta \theta_k (\omega)$, where $\Lambda_k$ is the thermo-elastic coupling, defined as:

\begin{equation}
  \Lambda^\theta_k = \int {\mathbf{S}}^x(\vec{r}) \, \mkern1mu{:} \, (\mathbf{c} \, \mkern1mu{:} \, \bm \alpha) \, \tilde{T}_k(\vec{r}) \, dV,
  \label{eq:Lambdak}
\end{equation}
where ${\mathbf{S}}^x(\vec{r})$ is the strain profile of the mechanical mode of interest, $\bm c$ is the stiffness tensor, $\bm \alpha$ is the thermal expansion tensor, and $\tilde{T}_k(\vec{r})$ is the temperature profile of the $k$-th thermal mode. In \autoref{fig:BackactionLoop}b, c, and d we present, respectively, schematic depictions of $\tilde{T}_k(\vec{r})$,  ${\mathbf{S}}^x(\vec{r})$, and the integrand of \autoref{eq:Lambdak}.

Summing over all thermal modes, the photothermal force can be written as:

\begin{equation}
  F_\mathrm{pt}(\omega) = \hbar G_x G_\mathrm{pt}(\omega) \Psi(\omega,\Delta) \delta x(\omega) n_{ph},
  \label{eq:PhotothermalBackaction}
\end{equation}
where $G_\mathrm{pt}(\omega)  = \omega_0 \kappa_\text{abs} \sum_k \left( \Lambda^\theta_k \frac{ R^\theta_k \chi^\theta_k(\omega)}{\tau_k} \right)$ is the photothermal equivalent to $G_x$, with one fundamental difference: $G_\mathrm{pt}(\omega)$ carries an extra phase delay. Applying \autoref{eq:Lambdak} to the definition of $G_\mathrm{pt}(\omega)$ and performing the sum over all thermal modes, we can define it in terms of the profile of thermal fluctuations, $\delta{T}(\vec{r},\omega) = \sum_k \left( \tilde{T}_k(\vec{r}) \frac{ R^\theta_k \chi^\theta_k(\omega)}{\tau_k} \right)$:

\begin{equation}
  G_\mathrm{pt}(\omega)  = \omega_0 \kappa_\mathrm{abs} \int {\mathbf{S}}^x(\vec{r}) \, \mkern1mu{:} \, (\mathbf{c} \, \mkern1mu{:} \, \bm \alpha) \, \delta{T}(\vec{r},\omega) \, dV.
  \label{eq:Gpt}
\end{equation}

The sign of $F_\mathrm{om}(\omega)$ is uniquely given by $\Psi(\omega, \Delta)$, as it further depends only on $G_x^2$, which is always positive. This is not the case for $F_\mathrm{pt}(\omega)$, with its overall sign also affected by the relative sign between $G_\mathrm{pt}(\omega)$ and $G_x$. The ability to control the thermoelastic coupling $\Lambda^\theta_k$, and consequentially $G_\mathrm{pt}(\omega)$, through the geometry of the device is precisely what we explore here to achieve deterministic control over the photothermal force: By controlling the overlap of the fields ${\mathbf{S}}^x(\vec{r})$ and $\delta T_k(\vec{r},\omega)$, the integral of \autoref{eq:Lambdak} can be drastically changed. Specifically, as we will see, both fields can have approximately orthogonal symmetries in realistic devices. A small breaking of symmetry through structural design can then lead to large changes of the backaction.

\subsection{Photothermal forces in zipper cavities}
\label{subsec:ForceZipper}

As shown by \autoref{eq:Gpt}, we can engineer either the thermal strain field, $ \bm\alpha \, \delta{T}(\vec{r},\omega)$ or the mechanical eigenmode strain field, $\mathbf{S}^x(\vec{r})$, to manipulate the photothermal force. Here we present an example of such engineering capability by breaking the symmetry of the mechanical strain field in a photonic crystal zipper cavity.

\begin{figure*}[htb!]
  \includegraphics[width = \textwidth]{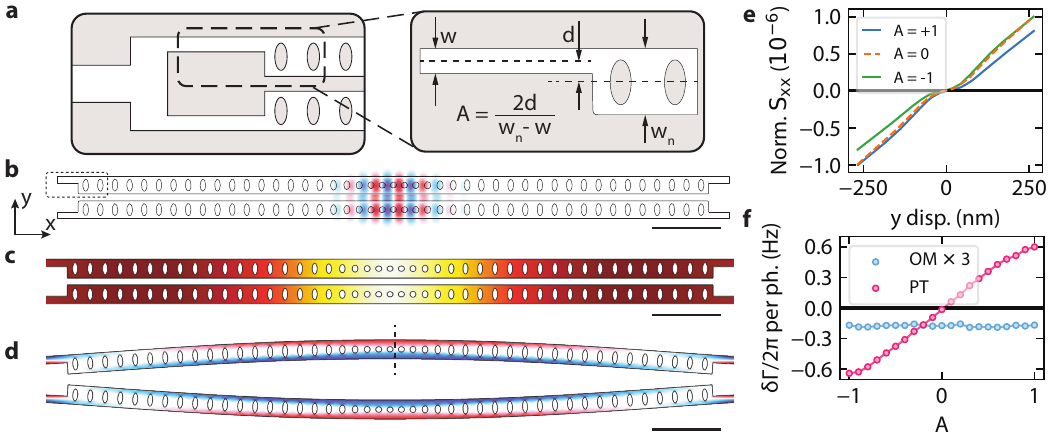}
  \caption{\small{\textbf{a} Definition of the tether asymmetry, $A$, in terms of other tether parameters. \textbf{b} Electric field in the $y$ direction for the even optical mode with frequency \si{196}{THz} with a \SI{2}{\um} scale indicated by the black line. \textbf{c} Imaginary part of the thermal fluctuations at the mechanical frequency. \textbf{d} $S_{xx}$ stress component for the odd mechanical mode with frequency 5.6 MHz. \textbf{e} Component $S_{xx}$ of the normalized strain  profile for different values of $A$ evaluated at the cut-line shown by the dashed line in \textbf{d}. \textbf{f} In red and blue we have, respectively, the expected maximal linewidth variation at the  blue side of the resonance due to photothermal and optomechanical effect, assuming typical experimental parameters for our devices and considering a temperature of \si{30}{K}, where the silicon expansion coefficient is negative. As the the optomechanical contributions are small we scaled it by 3 to improve the graph readability.}}
  \label{fig:zipper}
\end{figure*}

Zipper cavities are composed of two suspended nanobeams with elliptical holes, as shown in \autoref{fig:zipper}b. In a single beam, the optical modes are confined within the central region by a quasi-1D photonic crystal, in which a defect is introduced by changing the spacing and the size of the ellipses \cite{deotare_high_2009}. As the nanobeams are brought together, their modes hybridize, giving origin to even and odd supermodes \cite{deotare_coupled_2009, chan_optical_2009, eichenfield_picogram-_2009}. In our device, these are situated around \si{196}{THz}. As the gap decreases, the frequency of the even mode decreases, while the frequency of the odd mode increases.

Due to the supporting tethers, the flexural mechanical modes of the nanobeams also hybridize \cite{la2022nanomechanical}. The fundamental odd mode shown in \autoref{fig:zipper}b presents a very high $g_0$, due to its modulation of the gap between the nanobeams. 
For a gap of \si{200}{nm} $g_0/2\pi$ is around \si{-150}{kHz} for the even mode and \si{+30}{kHz} for the odd mode.

The optomechanical coupling arises from the overlap of the optical field and mechanical displacement, while the thermo-elastic one arises from the overlap of the thermal and mechanical strain fields, as shown in \autoref{eq:Gpt}. By changing the positioning of the supporting tethers, we can significantly affect the mechanical strain while keeping the optical and thermal fields constant. In our device, this can be achieved through the displacement of the tether with respect to the center line of the nanobeam, $d$, shown in \autoref{fig:zipper}a, with which we define the dimensionless tether asymmetry, $A = \frac{2d}{w - w_n}$.

In \autoref{fig:zipper}c the imaginary part of the temperature fluctuation profile, $\delta T(\vec{r},\Omega)$, is shown, as it is the part related to variations on the mechanical linewidth. Across the short transversal direction, heat is evenly spread, while across the longer longitudinal direction, it is mostly concentrated near the optical mode, as heat does not have enough time to homogenize during a mechanical period; $\Omega\tau_\mathrm{th}\gg1$. The mechanical strain profile, shown in \autoref{fig:zipper}d, is dominated by the $S_\mathrm{xx}$ component, which presents compressive stress in the inner part of the nanobeams and tensile stress in the outer part (for outwards displacement), with a neutral line close to the center of the nanobeam. For $A = 0$ the strain field has an almost perfect odd distribution in the $y$-direction with respect to the center line of a single nanobeam, as shown by the orange dashed line in \autoref{fig:zipper}e, while the temperature profile presents an even distribution leading to an odd integrand in \autoref{eq:Gpt} such that the photothermal coupling is approximately zero; $G_\textrm{pt}(\Omega) = 0$.

Crucially, by changing the asymmetry $A$ of the tether, we can control the intensity of the strain at the inner and outer edges of the nanobeam, as shown by the blue and green lines in \autoref{fig:zipper}e. The temperature profile remains almost completely unchanged, since it does not reach the tether region. This combination breaks the symmetry in \autoref{eq:Gpt}, allowing for the control of photothermal effects.

As the tether approaches one extremity, the strain increases at the opposite edge. For $A = 1$, the strain at the inner edge ($y<0$ in the graph) is larger in absolute terms than at the outer edge. This symmetry breaking mechanism is the dominant factor in the thermoelastic coupling, showing that by changing $A$ we can have a positive or negative photothermal force. In other words, if there is thermal expansion with $A = 1$, the inner edge will expand more, as the outer edge is constrained by the tethers. This will bend the device inward, bringing the two nanobeams closer together.

By controlling the sign of the thermo-elastic coupling we control the sign of the linewidth variation, as shown in \autoref{fig:zipper}f, where the calculated absolute maximal linewidth variation at the blue detuned side is shown as a function of $A$. As expected in the sideband non-resolved regime, the linewidth variation induced by optomechanical forces is small and does not depend on $A$, while the effects of photothermal forces, more or less proportional to $A$, are dominant, showing that changing the tether asymmetry allows tuning from amplification to cooling.

The length and width of the tethers are also relevant for the photothermal force. If the tethers are either too rigid or too compliant, the asymmetry effect is gradually lost, leading to small photothermal effects. As we increase the tether length, the mechanical frequency drops, and both optomechanical and photothermal effects fall. Nevertheless, the latter are less affected in such a way that the ratio between photothermal and optomechanical effects increases. In real devices, residual stress is another important aspect to consider as it can lead to large lateral warping of devices with asymmetric tethers, changing the intended gap between the nanobeams, and even leading to the collapse of the devices.

\subsection{Experimental verification}
\label{subsec:Experiment}

\begin{figure*}[htb!]
  \includegraphics[width = \textwidth]{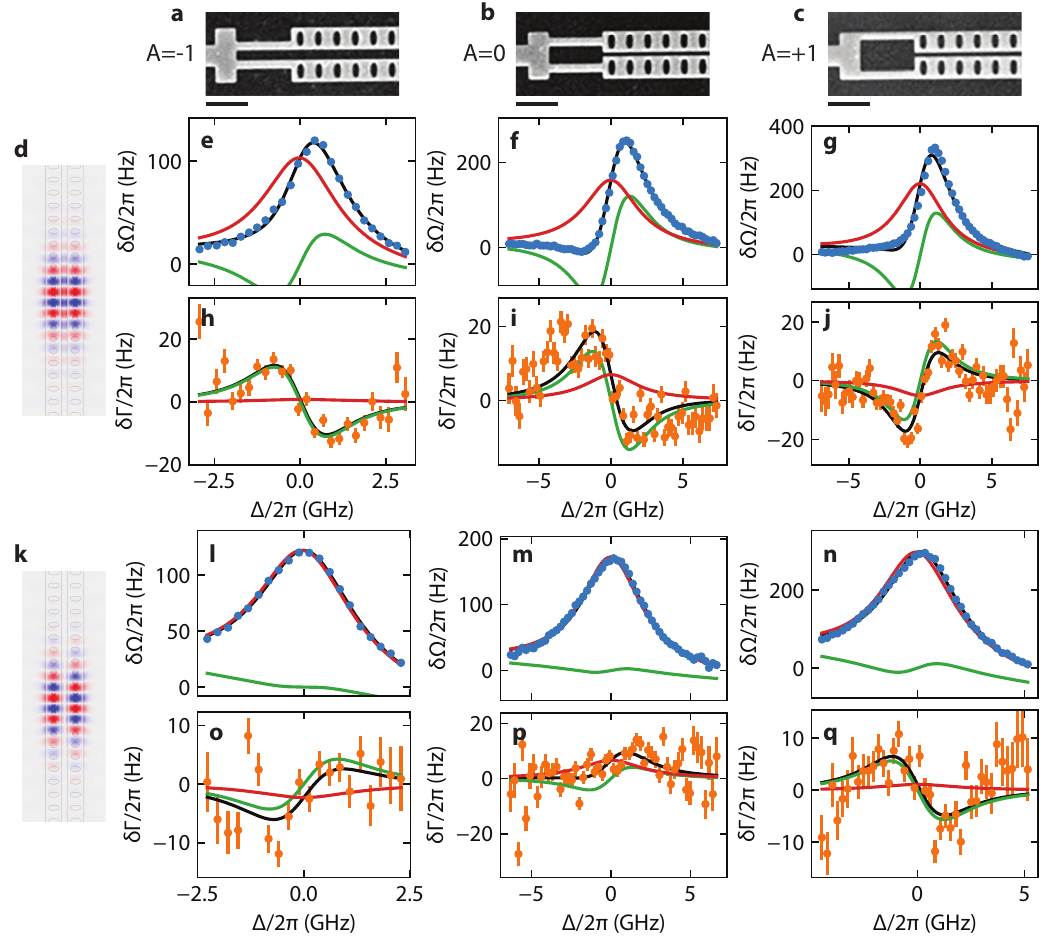}
  \caption{\small{\textbf{a}-\textbf{c} SEM micrographs of the tethers of three fabricated devices, with 1 µm scale bar. \textbf{d}, \textbf{k} Simulated $E_y$ mode profile of respectively the symmetric and anti-symmetric optical modes in the device. \textbf{e}-\textbf{g} (\textbf{h}-\textbf{j}) Measured frequency (linewidth) of the mechanical resonator as a function of optical detuning relative to the symmetric optical mode. Measured data is indicated by the blue (orange) dots with error bars. The black line is the complete fit to \autoref{eq:fitfunc}, with the red (green) line indicating the static (dynamic) contribution. \textbf{l}-\textbf{n} (\textbf{o}-\textbf{q}) Similar to \textbf{e}-\textbf{j}, but for the anti-symmetric optical mode.}}
  \label{fig:ExpResultRaw}
\end{figure*}

Using our model as a guide, we fabricate and characterize zipper cavities with varying asymmetrical tether positions, examples of which are shown in \autoref{fig:ExpResultRaw}a, b, and c. The details of the fabrication procedure are given in \ref{suppsec:Fabrication}. We characterize the devices in a closed-loop cryostat at \SI{30}{K} using optical heterodyne detection. Light is side-coupled to the nanocavities using an on-chip waveguide, that is interfaced to a lensed optical fiber at the side of the chip. The setup is described in \ref{suppsec:Setup}.

Simulations of the symmetric and anti-symmetric hybridized cavity modes are shown in \autoref{fig:ExpResultRaw}d and k, respectively. For each device, we step the laser wavelength over either of the optical resonances, and record a mechanical spectrum at each wavelength. From these spectra, the frequency and linewidth of the mechanical resonator can be determined through fitting the Lorentzian thermomechanical spectra. The results are shown in \autoref{fig:ExpResultRaw}e - j for the symmetric and \autoref{fig:ExpResultRaw}l - q for the anti-symmetric optical mode. The observed frequency and linewidth variations can be decomposed into static and dynamic contributions.

The dynamic contributions are related to the movement of the mechanical resonator (hence their nomenclature), and originate from the backaction loops discussed in \autoref{subsec:Backactions}. The static components, on the other hand, are purely due to the effects of the time-averaged local optical heating causing a change of the local temperature. The frequency of the mechanical mode increases due to the thermal contraction of the nanobeams --- the thermal expansion coefficient of silicon is negative at the temperatures used in our experiment\cite{vanhellemont_temperature_2014, middelmann_thermal_2015}. The linewidth, in turn, is altered due to temperature-dependent mechanical loss mechanisms\cite{mohanty_intrinsic_2002}.

The static components are therefore proportional to the local heating in the cavity, $\kappa_\mathrm{abs} n_\mathrm{ph}(\Delta)$, following an even squared Lorentzian function with respect to optical detuning $\Delta$. Consequently, static contributions should also not show a dependence on the tether asymmetry $A$ or optomechanical coupling $G_x$. This second assertion about $G_x$ can already be verified from \autoref{fig:ExpResultRaw}, by comparing the static contributions between the symmetric and anti-symmetric optical modes. These optical modes differ in $G_x$ by approximately a factor of 5, but show very similar magnitude static components.

The dynamic components, on the other hand, are related to dynamical backaction effects introduced in \autoref{subsec:Backactions} and are thus proportional to the derivative of the cavity Lorentzian; $\partial n_\mathrm{ph}/\partial \Delta$, which is odd with respect to $\Delta$. This leads to the following equations to fit to the experimental data:
\begin{subequations}
  \label{eq:fitfunc}
  \begin{align}
    \begin{split}
      \Omega &= \Omega_0 + \delta \Omega_\mathrm{drift} \cdot \Delta + \delta \Omega_\mathrm{dyn} \cdot \frac{32}{3 \sqrt{3}} \cdot \frac{\kappa^3 \cdot \Delta}{\left( 4 \Delta^2 + \kappa^2 \right)^2} \\ &\qquad + \delta \Omega_\mathrm{stat} \cdot \frac{\kappa^2}{4 \Delta^2 + \kappa^2}
    \end{split} \\
    \Gamma &= \Gamma_0 + \delta \Gamma_\mathrm{dyn} \cdot \frac{32}{3 \sqrt{3}} \cdot \frac{\kappa^3 \cdot \Delta}{\left( 4 \Delta^2 + \kappa^2 \right)^2} + \delta \Gamma_\mathrm{stat} \cdot \frac{\kappa^2}{4 \Delta^2 + \kappa^2},
    \label{eq:LWfit}
  \end{align}
\end{subequations}
where the various terms have been normalized such that their maxima are equal to $\delta\{\Omega,\Gamma\}_{\{\mathrm{dyn,stat}\}}$, similar to the value given for $\delta \Gamma_\mathrm{dyn}$ in \autoref{fig:zipper}f. The dynamic contributions can be related to the theory derived in \autoref{subsec:Backactions} through
\begin{subequations}
  \begin{align}
    \delta \Omega_\mathrm{dyn} &= \frac{1}{2 m_{\mathrm{eff}} \Omega } \mathrm{Re}\left[\Sigma_\mathrm{om}\left(\Omega, \frac{\kappa}{2\sqrt{3}}\right) + \Sigma_\mathrm{pt}\left(\Omega, \frac{\kappa}{2\sqrt{3}}\right)\right]   \\
    \delta \Gamma_\mathrm{dyn} &= \frac{-1}{m_{\mathrm{eff}} \Omega} \mathrm{Im}\left[\Sigma_\mathrm{om}\left(\Omega, \frac{\kappa}{2\sqrt{3}}\right) + \Sigma_\mathrm{pt}\left(\Omega, \frac{\kappa}{2\sqrt{3}}\right)\right],
    \label{eq:deltaGamma_dyn}
  \end{align}
\end{subequations}
with
\begin{subequations}
  \begin{align}
    \Sigma_\mathrm{om}(\omega, \Delta) =& \hbar (G_x)^2 \Psi(\omega,\Delta) n_\mathrm{ph}(\Delta), \\
    \Sigma_\mathrm{pt}(\omega, \Delta) =&  \hbar G_\mathrm{pt}(\omega) G_x \Psi(\omega,\Delta) n_\mathrm{ph}(\Delta).
  \end{align}
  \label{eq:inverse_suceptibilities}
\end{subequations}
The static contributions are limited to a phenomenological model as described above, but can still be used to infer information about effects that are proportional to temperature and time-averaged local heating.

As shown in \autoref{fig:zipper}e, the photothermal force is expected to be the main source of dynamic linewidth variation $\delta \Gamma_\mathrm{dyn}$ in these unresolved sideband systems. Indeed, our experimental results show values 20 times higher than what would be expected from pure optomechanics. Before delving deeper into quantitative analysis, we can already observe the tuning of the photothermal force from the opposite detuning dependence of the green lines in \autoref{fig:ExpResultRaw}h and j and \autoref{fig:ExpResultRaw}o and q.

Using \autoref{eq:inverse_suceptibilities}, the ratio between the mechanical linewidth variation due to photothermal  and optomechanical forces is given by
\begin{align}
  \frac{\delta \Gamma_\mathrm{dyn, pt}}{\delta \Gamma_\mathrm{dyn, om}} = \frac{\mathrm{Im}\left[ G_{pt}(\Omega)\right]}{G_x}  \frac{\mathrm{Re}\left[ \Psi \right]}{\mathrm{Im}\left[\Psi \right]} + \frac{\mathrm{Re}\left[ G_{pt}(\Omega)\right]}{G_x}.
\end{align}
In a sideband-unresolved mechanical resonator $\mathrm{Re}\left[ \Psi \right]/\mathrm{Im}\left[\Psi \right] \approx \kappa/4\Omega \gg 1$. Recalling the definition of $G_{\mathrm{pt},k}(\Omega) = \omega_0 \kappa_\text{abs} R^\theta_k\chi^{\theta}_k (\Omega)\Lambda^{\theta}_k / \tau_k$ we can examine the ratio between its real and imaginary parts. Using the fact that $R^\theta_k$ is proportional to $\tau_k$ and the assumption $1/\tau_k\ll\Omega$, which is true for all relevant thermal modes, we get $G_{\mathrm{pt},k}(\Omega) \propto (\Omega^2\tau_k)^{-1} + i/\Omega$. From this we see that $\mathrm{Im}\left[G_{\mathrm{pt},k}(\Omega)\right]/\text{Re}\left[G_{\mathrm{pt},k}(\Omega)\right] \approx \Omega \tau_k \gg 1$.

For the sake of our discussion here we can lump all the relevant thermal modes using an effective thermal response time such that $\mathrm{Im}\left[G_{\mathrm{pt}}(\Omega)\right]/\text{Re}\left[G_{\mathrm{pt}}(\Omega)\right] \approx \Omega \tau_{\mathrm{eff}} \gg 1$. Using $\kappa/2\pi  \approx \si{4}$~GHz and $\Omega/2\pi \approx \si{5}$~MHz we arrive at $\mathrm{Im}\left[ G_\mathrm{pt}(\Omega)\right] / G_x \sim 0.1$. Using these values we can show that $(\mathrm{Im} \left[ G_\mathrm{pt}(\Omega)\right]/G_x) \cdot (\mathrm{Re}\left[ \Psi \right] / \mathrm{Im}\left[\Psi \right]) \sim 20$, illustrating that the linewidth variation is dominated by the photothermal effect.

This result can also be used to estimate the ratio between the photothermal and optomechanical contributions to the mechanical frequency variation, which is given by
\begin{align}
  \frac{\delta \Omega_\mathrm{dyn, pt}}{\delta \Omega_\mathrm{dyn, om}} =& \frac{\mathrm{Re}\left[ G_{pt}(\Omega)\right]}{G_x} - \frac{\mathrm{Im}\left[ G_{pt}(\Omega)\right]}{G_x} \frac{\mathrm{Im}\left[ \Psi \right]}{\mathrm{Re}\left[\Psi \right]}\\
  \approx& \frac{\mathrm{Im}\left[ G_{pt}(\Omega)\right]}{G_x} \left( \frac{1}{\Omega\tau_{\mathrm{eff}}} - \frac{4\Omega}{\kappa} \right).
\end{align}
As $\mathrm{Im}\left[ G_{pt}(\Omega)\right]/G_x \sim 0.1$ and  $\left( (\Omega\tau_{\mathrm{eff}})^{-1} - 4\Omega/\kappa \right) \ll 1$ we can conclude that $\delta \Omega_\mathrm{dyn, pt}/\delta \Omega_\mathrm{dyn, om} \ll 1$, confirming that the mechanical frequency variation is indeed dominated by optomechanical effects.

\begin{figure*}[htb!]
  \includegraphics[width = \textwidth]{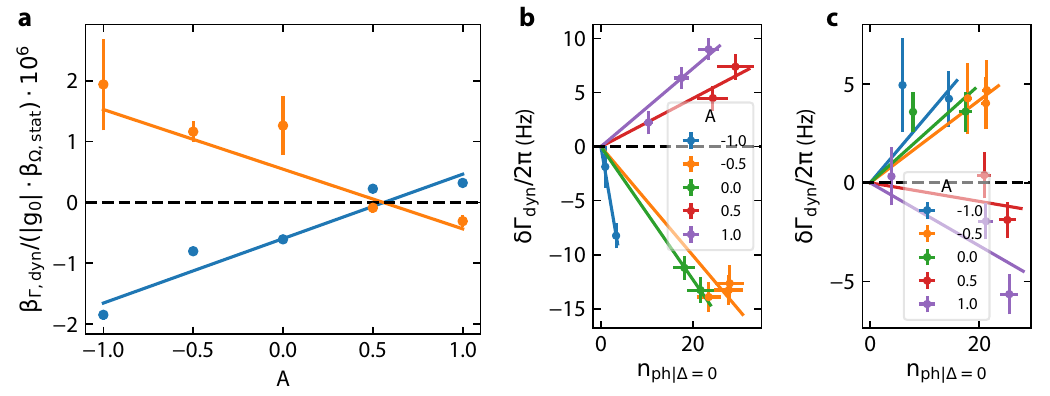}
  \caption{\textbf{a} The slope of the dynamic linewidth variation, $\beta_{\Gamma,\mathrm{dyn}} = \partial \delta \Gamma_\mathrm{dyn} / \partial n_\mathrm{ph}$, normalized by $|g_0|$ and $\beta_{\Omega,\mathrm{stat}}$, with a linear fit to the data for the symmetric (blue) and anti-symmetric (orange) optical modes. \textbf{b} (\textbf{c}) Measured dynamic linewidth variation for devices with various $A$ for the symmetric (anti-symmetric) optical mode. Lines are linear fits to the data using ODR.}
  \label{fig:ExpResultNormalized}
\end{figure*}

To allow a quantitative comparison of different physical devices, we repeat the procedure of sweeping the laser frequency over the optical resonances at various input powers. The slope $\beta_{\Gamma,\mathrm{dyn}} = \partial \delta \Gamma_\mathrm{dyn} / \partial n_{\mathrm{ph}|\Delta=0}$ of the dynamic linewidth variation $\delta \Gamma_\mathrm{dyn}$ to the number of photons at zero detuning $n_{\mathrm{ph}|\Delta=0}$ will allow us to compare $G_\mathrm{pt}(\Omega)$. The $\delta \Gamma_\mathrm{dyn}$ acquired from fitting to \autoref{eq:LWfit} is shown as a function of the number of photons at zero detuning $n_{\mathrm{ph}|\Delta=0}$ for the symmetric optical mode in \autoref{fig:ExpResultNormalized}b and for the anti-symmetric optical mode in \autoref{fig:ExpResultNormalized}c, together with linear fits, performed using orthogonal distance regression (ODR). The other components and their fits are shown in in \ref{suppfig:DataAnalysis}.

Importantly, we must correct for the variations in $G_x$ and $\kappa_\mathrm{abs}$ between devices. To enable this, we can extract information about $g_0$ and $\kappa_\mathrm{abs}$ from the derivative of the other frequency and linewidth variation components to the number of photons in the cavity.
In order to determine $g_0 = G_x x_\mathrm{zpf}$, we assume that the dynamic frequency change $\delta \Omega_\mathrm{dyn}$, plotted as the green line in \autoref{fig:ExpResultRaw}e-g and l-n, is purely optomechanical. This is reasonable considering the combination of optomechanical and photothermal ratios derived before. The $g_0$ of each optical mode can then be extracted from the slope $\beta_{\Omega,dyn} = \partial \delta \Omega_\mathrm{dyn} / \partial n_{\mathrm{ph}|\Delta=0}$ of the frequency variation $\delta \Omega_\mathrm{dyn}$ to the number of photons at zero detuning $n_{\mathrm{ph}|\Delta=0}$, with $|g_0| = \sqrt{ \frac{4}{3 \sqrt{3}} \beta_{\Omega,dyn} \kappa }$, as detailed in \ref{suppsec:DataAnalysis} and shown in \ref{suppfig:DataAnalysis}e and f. As we show in \ref{supptab:OptomechParams}, we obtain reasonable agreement between experiment and simulation.
Determining $\kappa_\mathrm{abs}$ is generally difficult, as the temperature cannot be measured directly in our devices. However, for small temperature changes $\delta T$, the static frequency variation $\delta \Omega_\mathrm{stat}$ is linearly proportional to $\delta T$  \cite{schmid_fundamentals_2023}, which in turn is linearly proportional to the product $\kappa_\mathrm{abs} n_\mathrm{ph}$ \cite{primo_accurate_2021}. Therefore, we can use the slope $\beta_{\Omega,\mathrm{stat}} = \partial \delta \Omega_\mathrm{stat} / \partial n_{\mathrm{ph}|\Delta=0}$, shown in \ref{suppfig:DataAnalysis}g and h, as a proxy for $\kappa_\mathrm{abs}$.

Finally, we can normalize the slope of the dynamic linewidth variation, $\beta_{\Gamma,\mathrm{dyn}}$ by $|g_0|$ and $\beta_{\Omega,\mathrm{stat}}$ and plot the result for all characterized geometries, as shown in \autoref{fig:ExpResultNormalized}a. This normalized value is proportional to the thermo-elastic force $F_\mathrm{pt}$, with its sign given by the relative sign between $G_\mathrm{pt}$ and $g_0$.

Looking at the change in normalized $\beta_{\Gamma,\mathrm{dyn}}$ as a function of $A$, we see that the sign and magnitude of the variation in linewidth can be controlled by changing the tether asymmetry, as predicted in \autoref{subsec:ForceZipper}. Comparing the results for even and odd optical modes, we see that the sign of the effect changes due to the change in the sign of $g_0$, while the magnitude of the effect remains the same, indicating that our normalized metric is indeed independent of $g_0$, as expected from photothermal effects.

Despite the clear qualitative agreement with the theory, we should note that according to our theory the $A = 0$ device is expected to have a very small linewidth variation, while this occurs in our experiments for slightly larger asymmetry $A\approx0.5$. Because the non-zero photothermal coupling originates from breaking the symmetry of the mechanical strain profile, this indicates that some other mechanism is inducing additional symmetry breaking.

Possible explanations could be an asymmetry in the sidewall properties between the inner and outer edges of the nanobeams due to different etch conditions in the gap. These could be the angle of the sidewall or the roughness of the surface. Assuming that we are close to the transition between ballistic and diffusive heat transport, even differences in roughness of the walls could lead to different effective temperatures. Finally, geometric asymmetry could also play a role, arising from a difference in alignment of the small asymmetric tethers with the long tethers to the substrate.

\vspace*{-4mm}
\subsection{Conclusion}
In summary, we showed that photothermal forces in nanoscale optomechanical devices can be controlled through strain engineering. We demonstrated this capability using the model system of a nanobeam zipper cavity. Using a theoretical model, we observed that the tether asymmetry strongly influences the strain profile across the nanobeams and thereby the thermoelastic overlap integral, while leaving the optomechanical coupling rate and mechanical frequency largely unaffected. Photothermal backaction is seen to depend very sensitively on geometrical parameters. For the same reason, we observed deviations due to inherent symmetry breaking, but geometrical control was seen to exceed that effect.

The ability to engineer photothermal forces enables a new degree of freedom in cavity optomechanics, allowing either suppression or enhancement of photothermal backaction. It could be used to suppress linewidth variation, thereby avoiding unwanted instabilities. Moreover, photothermal backaction could be maximized together with radiation pressure in optomechanical devices, leading to enhanced cooling and amplification performance in devices where purely optomechanical backaction is limited. This could allow the study of chaos in self-oscillating systems at lower input powers, or reaching stronger nonlinear regimes \cite{navarro-urrios_nonlinear_2017}. This enhancement of photothermal forces could also open up the exploration of macroscopic quantum mechanics with more massive mechanical resonators, which are usually deeply sideband-unresolved. It could allow studying the potential of photothermal forces for ground-state cooling in such systems, in regimes where regular optomechanical backaction cooling is  limited by the Doppler limit \cite{restrepo_classical_2011}.

\vspace*{-3mm}
\section*{Acknowledgement}
The authors thank Thiago P. Mayer Alegre for insightful discussions.

This work is part of the research programme of the Netherlands Organisation for Scientific Research (NWO). It is supported by an NWO Vrij Programma (680-92-18-04) grant. It is furthermore funded by the European Union, supported by ERC Grant 101088055 (Q-MEME). Views and opinions expressed are however those of the authors only and do not necessarily reflect those of the European Union or the European Research Council Executive Agency. Neither the European Union nor the granting authority can be held responsible for them.

C.M.K. acknowledges support by: São Paulo Research Foundation (FAPESP) through grants 18/15580-6, 18/25339-4, 20/06348-2, 22/14273-8; Coordenação de Aperfeiçoamento de Pessoal de Nível Superior - Brasil (CAPES) (Finance Code 001); and Financiadora de Estudos e Projetos (Finep).

\newpage
\appendix
\section*{Supplementary Information}
\renewcommand{\thesubsection}{Supplementary Section \arabic{subsection}}
\setcounter{section}{0}
\setcounter{subsection}{0}
\renewcommand{\figurename}{}
\renewcommand{\thefigure}{Supplementary Figure \arabic{figure}}
\setcounter{figure}{0}
\renewcommand{\tablename}{}
\renewcommand{\thetable}{Supplementary Table \arabic{table}}
\setcounter{table}{0}

\subsection{Fabrication}
\label{suppsec:Fabrication}

\begin{figure}[h!]
  \includegraphics[width=\columnwidth]{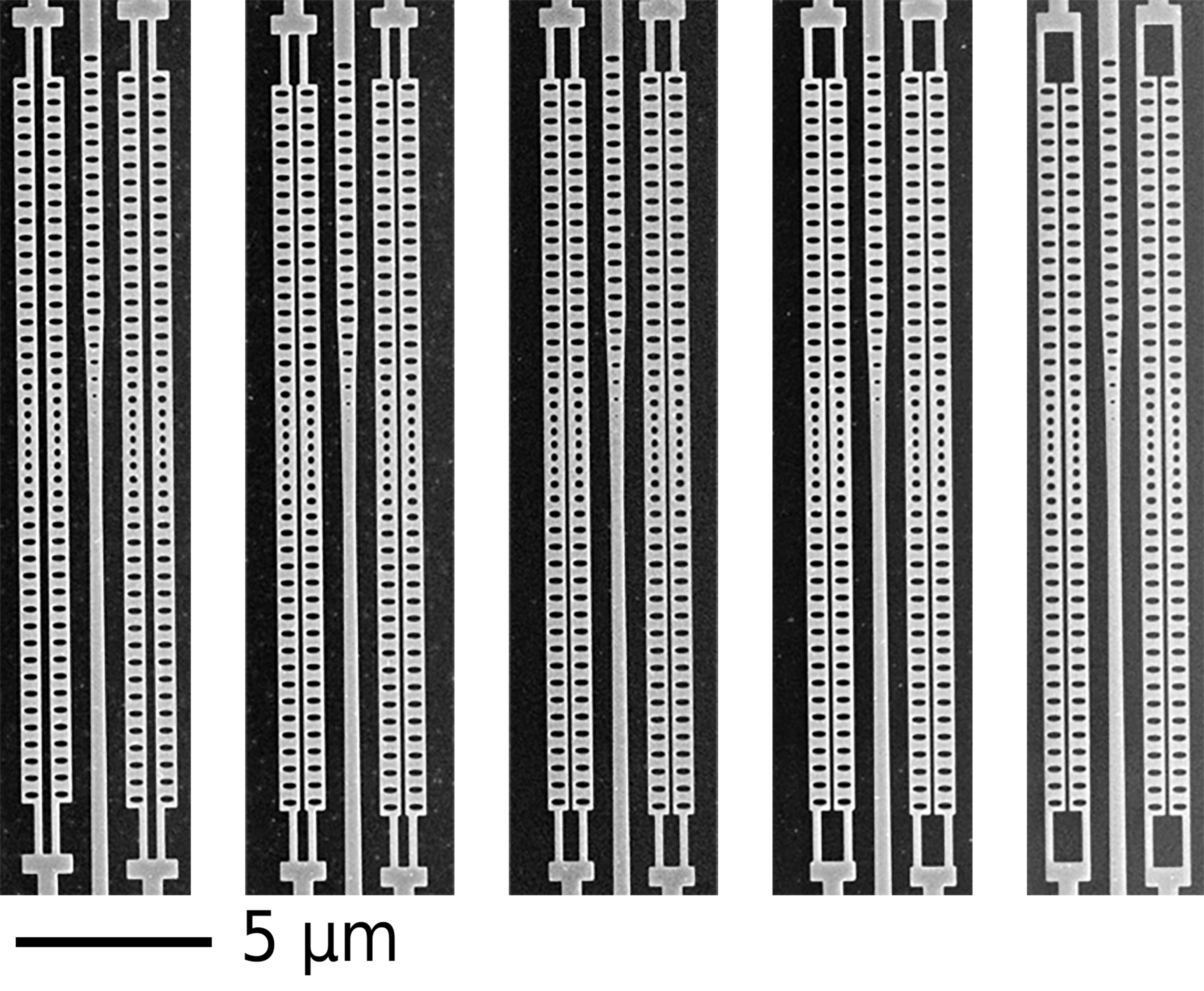}
  \caption{Overview of used devices. The asymmetry parameter $A$ varies from -1 to +1 from left to right in steps of 0.5.}
  \label{suppfig:Fabrication}
\end{figure}

The devices are fabricated using a standard E-beam lithography (EBL) process on a silicon-on-insulator (SOI) platform. First, a wafer with a 220 nm Si device layer, on top of 3 µm buried oxide and a thick Si substrate, is diced into 12x20 mm chips. The chip is cleaned with base piranha ($\mathrm{H_2 O / 30\% NH_4 OH/ 30\% H_2 O_2}$ 5:1:1 mixture). After this, a 100 nm layer of hydrogen silsesquioxane (HSQ) is spincoated on top. The chip is then cleaved to provide an edge with flat resist to which the devices will be aligned. This makes them accessible with a lensed fiber. The device geometry is then patterned in the resist aligned to the cleaved edge using a Raith Voyager EBL system, followed by development. This pattern is then etched into the Si device layer using Inductively Coupled Plasma Reactive Ion Etching (ICP RIE) in an Oxford Instruments PlasmaPro 100 Cobra with HBr/O$_2$ chemistry. Finally, the devices are underetched by hydrofluoric acid (HF) vapor etching using an SPTS µEtch. This simultaneously removes the leftover HSQ resist. \ref{suppfig:Fabrication} shows an overview of resulting devices with various asymmetries. Two zipper cavities (with slightly different scaling and thus optical frequencies) are positioned next to each feed waveguide to increase the number of devices that can be studied.

One aspect that can be seen in \ref{suppfig:Fabrication} is that for negative asymmetry we generally use devices with larger gaps separating the two beams of the zipper cavity than for positive asymmetry. This is because of residual stress in the SOI wafers that we encountered in earlier fabrication runs. Due to asymmetric attachment, the tensile stress would pull the two beams of negative asymmetry devices toward each other, causing the beams of devices with gaps below 150 nm to collapse. For positive asymmetry devices, smaller gaps can be realized. While this variation leads to variations in optomechanical coupling rate $g_0$, we compensate for this in the analysis as described in the main text. We note that the amount of tensile stress depends on the location of the chip in the wafer, which means that the impact of stress can vary from chip to chip.

\subsection{Device parameters}
\ref{supptab:MechParams} lists the mechanical frequencies and linewidths of the devices used. \ref{supptab:OptParams} lists the optical frequencies and linewidths of the cavity modes used. The uncertainty in the resonance frequencies is not listed, as it is on the order of ten Hz for the mechanical modes and MHz for the optical modes. The used gap sizes, along with simulated and measured values of $g_0$ for both modes, are listed in \ref{supptab:OptomechParams}.

\begin{table}[h!]
  \begin{tabular}{|l|l|l|}
    \hline
    A      & $\Omega/2\pi$ (MHz) & $\Gamma/2\pi$ (Hz) \\ \hline
    -1     & 5.767               & 144 $\pm$ 4          \\ \hline
    -0.5   & 5.752               & 160 $\pm$ 3          \\ \hline
    0      & 5.646               & 140 $\pm$ 2          \\ \hline
    0.5    & 5.760               & 142 $\pm$ 2          \\ \hline
    1      & 5.896               & 144 $\pm$ 2          \\ \hline
  \end{tabular}
  \caption{Mechanical parameters of the devices used in this work.}
  \label{supptab:MechParams}
\end{table}

\begin{table}[h!]
  \begin{tabular}{|l|l|l|l|}
    \hline
    A                     & Opt. symm. & $\omega/2\pi$ (THz) & $\kappa/2\pi$ (GHz) \\ \hline
    \multirow{2}{*}{-1}   & Even       & 192.027             & 3.824 $\pm$ 0.132   \\ \cline{2-4}
    & Odd        & 192.495             & 2.795 $\pm$ 0.009   \\ \hline
    \multirow{2}{*}{-0.5} & Even       & 186.933             & 3.344 $\pm$ 0.095   \\ \cline{2-4}
    & Odd        & 188.184             & 4.294 $\pm$ 0.074   \\ \hline
    \multirow{2}{*}{0}    & Even       & 187.360             & 4.525 $\pm$ 0.029   \\ \cline{2-4}
    & Odd        & 188.308             & 4.686 $\pm$ 0.043   \\ \hline
    \multirow{2}{*}{0.5}  & Even       & 189.875             & 3.669 $\pm$ 0.003   \\ \cline{2-4}
    & Odd        & 190.760             & 4.271 $\pm$ 0.021   \\ \hline
    \multirow{2}{*}{1}    & Even       & 189.848             & 3.899 $\pm$ 0.003   \\ \cline{2-4}
    & Odd        & 190.673             & 4.277 $\pm$ 0.011   \\ \hline
  \end{tabular}
  \caption{Optical parameters of the devices used in this work.}
  \label{supptab:OptParams}
\end{table}

\begin{table}[]
  \begin{tabular}{|l|l|l|l|l|}
    \hline
    A                     & Gap (nm)             & Opt. symm. &
    \begin{tabular}[c]{@{}l@{}}Simulated\\ $g_0/2\pi$ (kHz)
    \end{tabular} &
    \begin{tabular}[c]{@{}l@{}}Measured\\ $|g_0|/2\pi$ (kHz)
    \end{tabular} \\ \hline
    \multirow{2}{*}{-1}   & \multirow{2}{*}{260} & Even       & -69.5                                                                & 104.2 $\pm$ 2.4                                                       \\ \cline{3-5}
    &                      & Odd        & 38.4                                                                 & 19.8 $\pm$ 0.4                                                        \\ \hline
    \multirow{2}{*}{-0.5} & \multirow{2}{*}{175} & Even       & -190.1                                                               & 164.1 $\pm$ 2.5                                                       \\ \cline{3-5}
    &                      & Odd        & 50.5                                                                 & 20.4 $\pm$ 2.7                                                        \\ \hline
    \multirow{2}{*}{0}    & \multirow{2}{*}{200} & Even       & -143.7                                                               & 140.1 $\pm$ 0.4                                                       \\ \cline{3-5}
    &                      & Odd        & 45.7                                                                 & 24.2 $\pm$ 0.1                                                        \\ \hline
    \multirow{2}{*}{0.5}  & \multirow{2}{*}{200} & Even       & -132.9                                                               & 128.0 $\pm$ 0.9                                                       \\ \cline{3-5}
    &                      & Odd        & 54.5                                                                 & 46.7 $\pm$ 2.5                                                        \\ \hline
    \multirow{2}{*}{1}    & \multirow{2}{*}{200} & Even       & -134.1                                                               & 120.0 $\pm$ 0.4                                                       \\ \cline{3-5}
    &                      & Odd        & 49.8                                                                 & 46.0 $\pm$ 3.5                                                        \\ \hline
  \end{tabular}
  \caption{Optomechanical parameters of the devices used in this work.}
  \label{supptab:OptomechParams}
\end{table}

\subsection{Setup}
\label{suppsec:Setup}
\begin{figure*}[htb!]
  \includegraphics[width=0.7\textwidth]{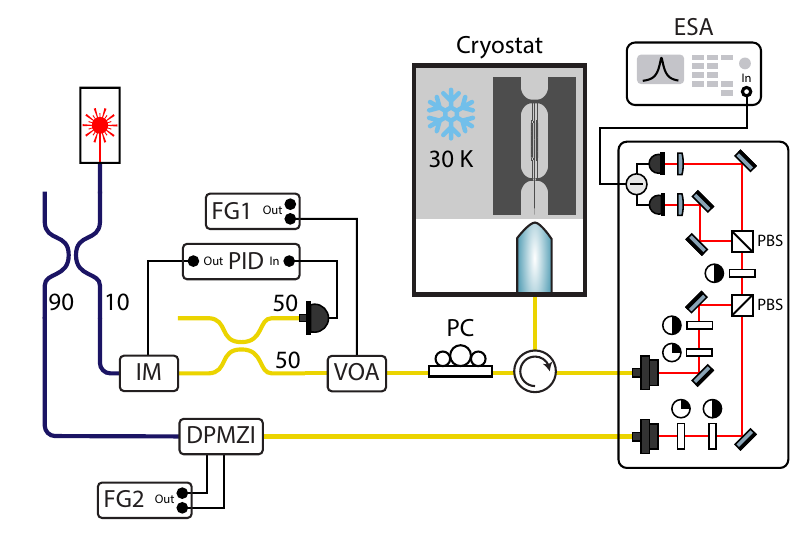}
  \caption{Experimental setup. PID: PID controller. FG: Function generator. IM: Intensity modulator. VOA: Variable optical attenuator. PC: Polarization controller. DPMZI: Dual parallel phase-modulated Mach-Zehnder interferometer. ESA: Electronic spectrum analyzer.}
  \label{suppfig:Setup}
\end{figure*}

The characterization of all devices was performed using the setup depicted in \ref{suppfig:Setup}. The light source is a Toptica CTL 1550 tunable laser with a wavelength range from 1510 to 1630 nm. A 90:10 fiber beamsplitter splits the laser light into a local oscillator (LO) and signal arm. The intensity of the light in the signal arm is stabilized using a PID controller. With a subsequent variable optical attenuator (VOA), the power can be reduced electronically to enable power sweeps. This light is then routed through a circulator and coupled into the device using a lensed fiber. The same lensed fiber collects the reflected light, which, depending on detuning, has entered the cavity and has had the mechanical motion imprinted on its phase. The local oscillator arm goes through a dual parallel phase-modulated Mach-Zehnder interferometer (DPMZI), which displaces the frequency of the light by 40 MHz by single-sideband modulation. The light from both arms is coupled into free space, where their polarizations are rotated such that they are orthogonal before passing through the first polarizing beamsplitter (PBS), and then rotated by $45^\circ$ before passing through the second PBS. That PBS splits the light fields such that they can interfere on both arms of the balanced detector. The electronic spectrum analyzer (ESA) then takes the Fourier transform of the differential photocurrent, giving us our mechanical spectra.

\subsection{Data Analysis}
\label{suppsec:DataAnalysis}
\begin{figure*}[htb!]
  \includegraphics[width=\textwidth]{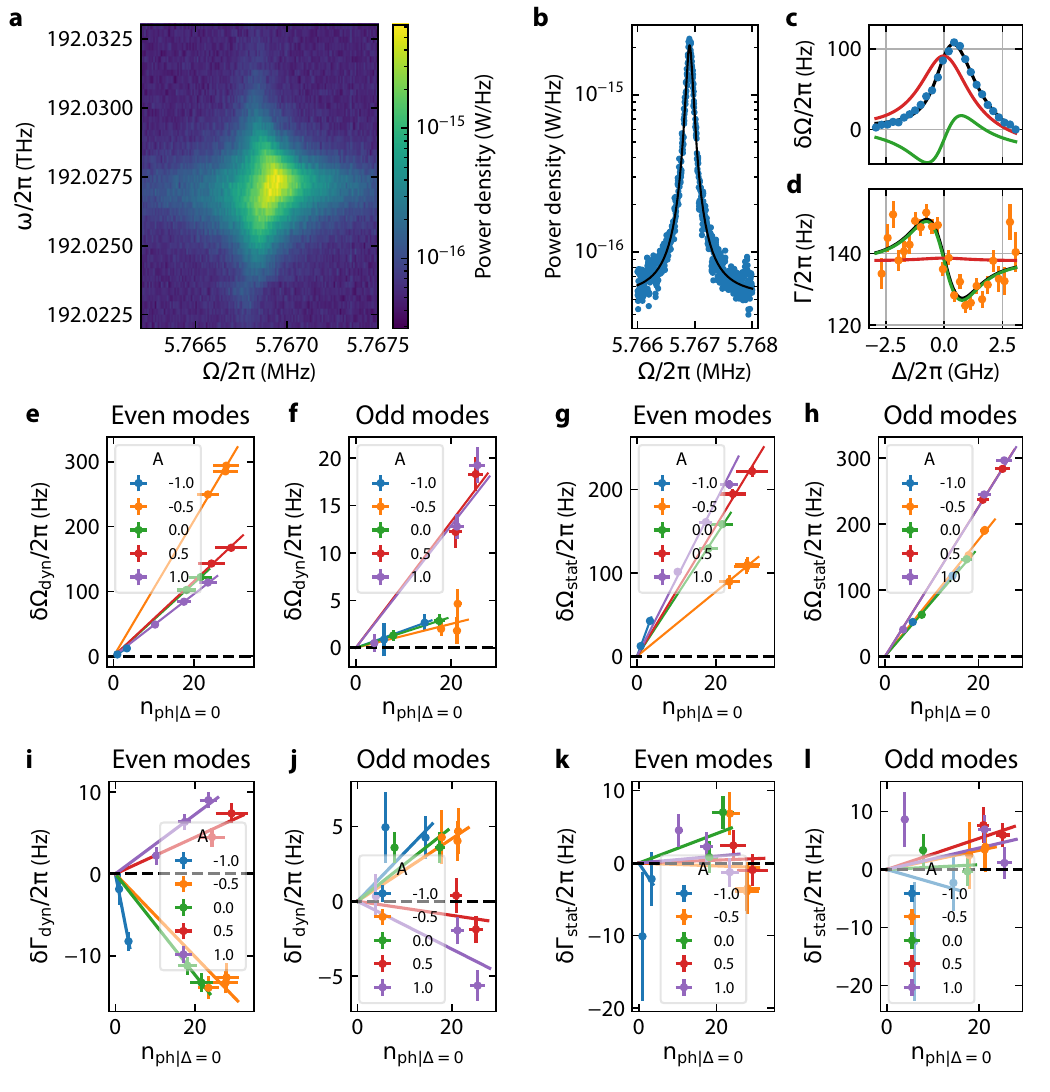}
  \caption{\textbf{a} Optical frequency sweep of the device with $A=-1$. \textbf{b} Mechanical spectrum at one optical frequency. \textbf{c}, \textbf{d} Extracted mechanical frequency and linewidth as function of optical detuning, with symmetric static (red) and anti-symmetric dynamic (green) components of the fit (black). \textbf{e}-\textbf{l} Extracted dynamic and static frequency and linewidth contributions for all asymmetries $A$ as a function of peak cavity photon number.}
  \label{suppfig:DataAnalysis}
\end{figure*}

In this subsection, we will describe the data analysis steps that yield the final normalized linewidth variation $\beta_{\Gamma,\mathrm{dyn}}/(|g_0|\beta_{\Omega,\mathrm{stat}})$ that is plotted in \autoref{fig:ExpResultNormalized}a and discussed in the main text. The different steps are shown in the panels of \ref{suppfig:DataAnalysis}. Firstly, we perform a wavelength sweep over the optical mode at fixed input power, recording the mechanical spectrum at each wavelength. An example of such a sweep for the $A=-1$ device is shown in \ref{suppfig:DataAnalysis}a. At each wavelength, the mechanical frequency and linewidth are determined using a fit, as shown in \ref{suppfig:DataAnalysis}b. Plotting the frequency (linewidth) at each detuning yields the blue (orange) data in \ref{suppfig:DataAnalysis}c (d), with the error bars determined from the covariance matrix of the least squares fit. We can now fit this data with a symmetric and anti-symmetric component, being the optical Lorentzian response and derivative of this Lorentzian, for the static and dynamic contributions, respectively:

\begin{align}
  \label{suppeq:fitfunc}
  \begin{split}
    \Omega &= \Omega_0 + \delta \Omega_\mathrm{drift} \cdot \Delta + \delta \Omega_\mathrm{dyn} \cdot \frac{32}{3 \sqrt{3}} \cdot \frac{\kappa^3 \cdot \Delta}{\left( 4 \Delta^2 + \kappa^2 \right)^2} \\
    & \qquad + \delta \Omega_\mathrm{stat} \cdot \frac{\kappa^2}{4 \Delta^2 + \kappa^2}
  \end{split} \\
  \Gamma &= \Gamma_0 + \delta \Gamma_\mathrm{dyn} \cdot \frac{32}{3 \sqrt{3}} \cdot \frac{\kappa^3 \cdot \Delta}{\left( 4 \Delta^2 + \kappa^2 \right)^2} + \delta \Gamma_\mathrm{stat} \cdot \frac{\kappa^2}{4 \Delta^2 + \kappa^2}.
\end{align}

Here, the prefactors originate from normalization to facilitate fitting directly to the experimental data, without requiring external quantities, allowing calibrations of powers and efficiencies to be done separately. As an example, in the bad cavity limit ($\Omega \ll \kappa$) the expression for the dynamical frequency variation, due to the optical spring effect, is given by
\begin{align}
  \delta \Omega_{m} (\Delta) \approx g_0^2 n_{cav}(\Delta) \frac{2 \Delta}{\Delta^2 + \kappa^2/4}.
\end{align}
We fill in $n_{cav}(\Delta) = \frac{P_{in}}{\hbar \omega_L} \frac{\kappa_{ex}}{\Delta^2 + \kappa^2 /4}$, solve for the maximum at $\frac{\partial \delta \Omega_{m}}{\partial \Delta} = 0$, and use the value found to normalize the $\delta \Omega_{dyn}$ term in \autoref{suppeq:fitfunc}.

An example of this fit is shown in \ref{suppfig:DataAnalysis}c and d for frequency and linewidth, respectively. The optical resonance frequency and linewidth are determined separately from a fit to the area of the mechanical peak, and kept fixed for these fits. The static contribution is plotted in red, and the dynamic contribution in green. Repetition of this procedure at multiple powers gives the data points that belong to one device in \ref{suppfig:DataAnalysis}e - l. The error bars represent standard errors obtained from the covariance of the fit parameters. From a linear fit to these points, we can then get the slopes $\beta_{\{ \Omega,\Gamma \},\{ dyn,stat \}}$.

The final quantity required for normalization in \autoref{fig:ExpResultNormalized}a is $g_0$, which we can calculate from $\beta_{\Omega,dyn}$ with $|g_0| = \sqrt{ \frac{4}{3 \sqrt{3}} \beta_{\Omega,dyn} \kappa }$.

\subsection{Theoretical model}
\label{suppsec:Theoretical model}

Here we extend the discussion on the mechanisms behind the photothermal backaction cycle discussed in the main text, by reviewing the model established in \cite{primo_accurate_2021}. The dynamics of the optical and mechanical modes are treated in the standard optomechanical way, with lumped element coupled mode theory equations, where the coupling rates are given by overlap integrals of the optical and mechanical fields \cite{wiederhecker2019brillouin}.

The first step in the modeling consists of knowing how an arbitrary oscillating temperature field drives the mechanical mode of interest, from \cite{primo_accurate_2021} we have
\begin{equation}
  m_\text{eff}\big(\delta\ddot{x}+\Gamma\delta\dot{x}+\Omega^2 \delta x\big) = \int \bm S^x(\vec{r}) \mkern1mu{:}\, \big( \bm c \mkern1mu{:} \bm S^\theta(\vec{r}, t) \big) \, dV,
  \label{eq:mech}
\end{equation}
where $\delta x$ represents the amplitude of fluctuations in the mechanical mode, $m_\text{eff}$, $\Gamma$, and $\Omega$ are its effective mass, decay rate, and angular frequency, $\bm c$ is the stiffness tensor of the medium, the operator “:” is the tensor contraction operation, $\bm S^x(\vec{r})$ is the normalized strain profile of the mechanical mode of interest and $ \bm S^\theta(\vec{r}, t)$ is the thermal strain profile, which is given by
\begin{equation}
  \bm S^\theta(\vec{r}, t) = \bm \alpha \delta T(\vec{r}, t),
\end{equation}
where $\bm \alpha$ is the thermal expansion tensor and $\delta T(\vec{r},t)$ is the temperature fluctuation profile.

In the case of the photothermal backaction $\delta T(\vec{r},t)$ is generated by the photon number fluctuations, following the equation for diffusive heat transport:
\begin{equation}
  c_p \rho \partial_t \delta T(\vec{r},t) =  \nabla \cdot \big(k_{\text{th}} \nabla \delta T(\vec{r},t)\big) + \dot{Q}(\vec{r},t),
  \label{eq:heat}
\end{equation}
where $c_p$, $\rho$ and $k_{\text{th}}$ are, respectively, the specific heat, the mass density, and the thermal conductivity of the medium. $\dot Q$ is the fluctuating heat source due to the optical absorption, assuming linear absorption at the volume. It is defined inside the region in space where absorption occurs, given by the heat source
\begin{equation}
  \dot{Q}(\vec{r},t) = \frac{\kappa_\text{abs} \, \varepsilon_r(\vec{r}) |\vec{e}(\vec{r})|^2}{\int_{V_\text{abs}} \varepsilon_r(\vec{r}) |\vec{e}(\vec{r})|^2 dV} \; \hbar \omega_0\left( a_0^* \delta a(t) + a_0 \delta a^*(t) \right),
  \label{eq:heat-source}
\end{equation}
where $a_0$ represents the mean photon amplitude, $\delta a(t)$ are the photon amplitude fluctuations, $\kappa_\text{abs}$ and $\omega_0$ are the optical absorption rate and angular optical frequency, $\varepsilon_r(\vec{r})$ is the medium dielectric constant, $\vec{e}(\vec{r})$ is the normalized optical electric field, and $\int_{\text{abs}}$ indicates integration over the region of the device where absorption occurs. The term $\left( a_0^* \delta a(t) + a_0 \delta a^*(t) \right)$ approximates the linearized photon number fluctuations $\delta n(t)$, assuming that the overall fluctuations are small compared to the mean photon number.

Following the derivation from Supplementary Section S2 of Primo \textit{et al.}\cite{primo_accurate_2021}, if we set the source of heat to zero, $\dot{Q}(\vec{r},t)=0$, \autoref{eq:heat} is separable with solutions of the form $\delta T(\vec{r}, t) = \delta \tilde{T}(\vec{r})e^{-t/\tau}$ giving an eigenvalue problem for the thermal modes:
\begin{align}
  \nabla \cdot\left(k_{\mathrm{th}} \nabla \delta \tilde{T}(\vec{r})\right)=-\frac{1}{\tau} c_p \rho \delta \tilde{T}(\vec{r}).
\end{align}

As both operators $\nabla \cdot\left(k_{\mathrm{th}} \nabla\right)$ and $c_p \rho$ in this equation are Hermitian and the latter is positive definite, this equation has a complete set of eigenmodes:

\begin{equation}
  \delta T(\vec{r},t) = \sum_k \delta \theta_k(t)  \tilde{T}_k(\vec{r}),
\end{equation}
where $\delta\theta_k(t)$ represents the amplitude of the $k$-th thermal mode and $\tilde{T}_k(\vec{r})$ represents its normalized profile.
Assuming fixed temperature or constant heat flow at the boundaries, it can be shown that these eigenmodes are orthogonal. The above simplifications of the heat equation are more commonly known as reducing it to a Sturm-Liouville problem, with associated theory\cite{guenther_sturm-liouville_2018}.

Using the properties of this complete set of eigenmodes and \autoref{eq:heat}, one can derive dynamic equations for the amplitude of each thermal mode:
\begin{equation}
  \delta\dot{\theta}_k = -\frac{1}{\tau_k}\delta\theta_k + \frac{\int{\dot{Q}(\vec{r},t) \tilde T_k(\vec{r})dV}}{\int{c_p\rho \tilde T_k^2(\vec{r}) dV }},
  \label{theta-evo-int}
\end{equation}
where $\tau_k$ is the thermal decay time for the $k$-th mode. Applying \autoref{eq:heat-source} on the right-hand side of \autoref{theta-evo-int} the dynamic equation can be written as:
\begin{equation}
  \delta\dot{\theta}_k = -\frac{1}{\tau_k}\delta\theta_k + \frac{\hbar \omega_0 \kappa_{\text{abs}} R^\theta_{k}}{\tau_k}  \left( a_0^* \delta a + a_0 \delta a^* \right),
  \label{nl_theta-evo}
\end{equation}
where the thermal resistance of the $k$-th mode is defined as:
\begin{equation}
  R^\theta_{k} = \tau_k \frac{\int_{V_\text{abs}}{\varepsilon_r |\vec{e}|^2 \tilde T_k dV}}{\int_{V_\text{abs}} \varepsilon_r |\vec{e}|^2 dV\int{c_p\rho \tilde T_k^2 dV}},
  \label{eq:thermal_resistance}
\end{equation}

The same modal expansion of the temperature field is performed on the right-hand side of \autoref{eq:mech}:
\begin{equation}
  \int \bm S^x(\vec{r}) \mkern1mu{:}\, \big( \bm c \mkern1mu{:} \bm S^\theta(\vec{r}, t) \big) \, dV = \Lambda_{k}^{\theta} \delta\theta_k(t),
\end{equation}
where $\Lambda^{\theta}_{k}$ is the thermo-elastic coupling between the $k$-th thermal mode and the mechanical mode:
\begin{equation}
  \Lambda^{\theta}_{k} = \int {\mathbf{S}}^x \, \mkern1mu{:} \, (\mathbf{c} \, \mkern1mu{:} \, \bm \alpha) \, \tilde{T}_k \, dV
\end{equation}

As such, the full set of lumped parameter equations for the coupled dynamics of the optical, mechanical, and thermal mode fluctuations is given by
\begin{align}
  \begin{split}
    \delta\dot{a}=&i
    G_{x} \delta x a_{0}+(i\Delta_0 -\frac{\kappa}{2})\delta a,\\
    \delta\ddot{x}+\Gamma\delta\dot{x}+\Omega^{2}\delta x =& \frac{\hbar G_{x}}{m_\text{eff}}(a_0^{*}\delta a+ a_0\delta a^{*})+\sum_k\frac{\Lambda^\theta_k}{m_\text{eff}}\delta\theta_k,\\
    \delta \dot{\theta}_k =& -\frac{1}{\tau_k}\delta\theta_k + \frac{\hbar \omega_0 \kappa_\text{abs}R^\theta_k}{\tau_k}(a_0^{*}\delta a+ a_0\delta a^{*}),
  \end{split}
\end{align}
where $\kappa$ is the total optical decay rate, $G_x$ is the optomechanical coupling\footnote{The normalized optomechanical coupling is given by $g_0 = G_\textrm{x}\sqrt{\frac{\hbar}{2m_\textrm{eff}\Omega}}$}, and $\Delta_0 = \omega_l - \omega_0$ is the detuning between the angular frequencies of the laser, $\omega_l$, and of the optical mode $\omega_0$.

These linearized equations can be solved in the frequency domain:
\begin{align}
  \begin{split}
    \left[(\Delta_0+ \omega) +i\frac{\kappa}{2}\right]\delta a(\omega)=& - G_{x}\delta x(\omega) a_{0},\\
    \begin{split}
      \big[(\Omega^{2}-\omega^2)-i\omega \Gamma\big] \delta x(\omega)=& \frac{\hbar G_{x}}{m_\text{eff}}\big[a_0^{*}\delta a(\omega)+ a_0[\delta a]^*(\omega)\big]\\
      &\qquad+\sum_k\frac{\Lambda^{\theta}_k}{m_\text{eff}} \delta\theta_k(\omega),
    \end{split}\\
    \left( -i\omega + \frac{1}{\tau_k} \right) \delta\theta_k(\omega) =& \frac{\hbar \omega_0 \kappa_\text{abs}R^\theta_k}{\tau_k}\big[a_0^{*}\delta a(\omega)+ a_0[\delta a]^*(\omega)\big].
  \end{split}
  \label{eq:dyn}
\end{align}
Using the fact that $\delta x(t)$ and $\delta \theta(t)$ are real variables, these equations can be solved in such a way that the equation for mechanical dynamics can be written as:
\begin{widetext}
  \begin{equation}
    m_\text{eff} \big[(\Omega^{2}-\omega^2)-i\omega \Gamma\big] \delta x(\omega)= - \left( \hbar G_{x} + \hbar\omega_0 \kappa_\text{abs} \sum_k \frac{\Lambda^{\theta}_kR^\theta_k\chi^{\theta}_k (\omega)}{\tau_k}\right)
    \Psi(\omega,\Delta_0)|a_0|^2 G_{x} \delta x(\omega)
    ,
    \label{eq:coupled_mech_dyn}
  \end{equation}
\end{widetext}
where
\begin{equation}
  \chi^{\theta}_k (\omega) = \frac{1}{1/\tau_k - i\omega}
  \label{eq:ThermalSusceptibility}
\end{equation}
is the thermal susceptibility of the $k$-th mode and
\begin{equation}
  \Psi (\omega, \Delta_0) = \frac{1}{(\Delta_0+ \omega) +i\frac{\kappa}{2}} + \frac{1}{(\Delta_0 - \omega) -i\frac{\kappa}{2}},
\end{equation}
is the symmetrized optical susceptibility.

The left-hand side of \autoref{eq:coupled_mech_dyn} is essentially the inverse of the bare mechanical susceptibility $\chi_{m}^{-1}(\omega)=m_{\mathrm{eff}}\left[\left(\Omega^{2}-\omega^{2}\right)-i \Gamma \omega\right]$. As such, the right-hand side, which is also proportional to $\delta x(\omega)$, can be seen as optomechanical ($\Sigma^\text{RP}(\omega, \Delta_0)$) and photothermal ($\Sigma^\theta(\omega, \Delta_0)$) contributions to a ``dressed'' inverse mechanical susceptibility:
\begin{equation}
  \chi_{m, \mathrm{eff}}^{-1}(\omega)=\chi_{m}^{-1}(\omega)+\Sigma^\text{RP}(\omega)+\Sigma^\theta(\omega),
\end{equation}
where
\begin{subequations}
  \begin{align}
    \Sigma^\text{RP}(\omega, \Delta_0) =& \hbar (G_{x})^2 \Psi(\omega,\Delta_0) |a_0|^2, \\
    \Sigma^\theta(\omega, \Delta_0) =&  \hbar \omega_0 \kappa_\text{abs} \sum_k \frac{ R^\theta_k \Lambda^\theta_k \chi^\theta_k(\omega)}{\tau_k} G_{x} \Psi(\omega,\Delta_0)|a_0|^2.
  \end{align}
\end{subequations}

In order to simplify our notation, it is useful to define here an ``effective photothermal coupling'' accounting for the multiple thermal modes, where all the calculated overlap integrals related to the photothermal interaction are lumped together:
\begin{equation}
  G_\mathrm{pt}(\omega)  = \omega_0 \kappa_\text{abs} \sum_k \frac{R^\theta_k\Lambda^{\theta}_k\chi^{\theta}_k (\omega)}{\tau_k}.
\end{equation}
This effective response can be used, for example, to define the equivalent single-photon photothermal force as $\hbar G_\mathrm{pt}(\omega)$ that can be more straightforwardly compared to the optomechanical single photon force $\hbar G_{x}$.

In the weak coupling regime discussed here, both $\Sigma^\text{RP}(\omega)$ and $\Sigma^\theta(\omega)$ are more or less constants over the regions where the mechanical susceptibility is relevant --- a band of size $\Gamma$ around the frequency $\Omega$. As such, the effect of the coupling is to change the mechanical frequency and linewidth:
\begin{subequations}
  \begin{align}
    \Omega_{\text{eff}} = \Omega + \delta \Omega \\
    \Gamma_{\text{eff}} = \Gamma + \delta \Gamma,
  \end{align}
\end{subequations}
where
\begin{subequations}
  \begin{align}
    \delta \Omega &= \frac{1}{2 m_{\text{eff}} \Omega } \text{Re}\left[\Sigma^\text{RP}(\Omega, \Delta_0) + \Sigma^\theta(\Omega, \Delta_0)\right]  , \\
    \delta \Gamma &= \frac{-1}{m_{\text{eff}} \Omega} \text{Im}\left[\Sigma^\text{RP}(\Omega, \Delta_0) + \Sigma^\theta(\Omega, \Delta_0)\right]  .
  \end{align}
\end{subequations}
The real and imaginary parts of the above expressions are related to the in-phase and in-quadrature responses of the backaction loops evaluated at frequency $\Omega$. The in-phase response generates a conservative force, which only affects the mechanical frequency, while the in-quadrature response is related to dissipative forces which affect the mechanical linewidth.

Separating optomechanical and photothermal contributions to the linewidth variation, we arrive at
\begin{subequations}
  \begin{align}
    \delta \Gamma^{\text{RP}} &= - \frac{\hbar (G_{x})^2 |a_0|^2}{ m_{\text{eff}} \Omega } \text{Im}\left[ \Psi(\Omega,\Delta_0) \right]    \\
    \delta \Gamma^\theta &= -\frac{\hbar G_{x} |a_0|^2 }{ m_{\text{eff}} \Omega } \text{Im}\left[G_{\mathrm{pt}}(\Omega)  \Psi(\Omega,\Delta_0)\right]
    \label{eq:delta_gamma_theta},
  \end{align}
\end{subequations}
For the optomechanical contribution only the delay given by the optical response is relevant, while for the photothermal contribution both the optical and thermal responses are present. Our devices operate in the unresolved sideband regime $\Omega\ll\kappa$, where the mechanical time scale is much slower than the optical time scale. In this regime, we have $\text{Im}\left[ \Psi(\Omega,\Delta_0) \right] \ll \text{Re}\left[ \Psi(\Omega,\Delta_0) \right]$, and as such $\delta \Gamma^{\text{RP}}$ is much smaller than $\delta \Omega^{\text{RP}}$.

Comparison of the mechanical and thermal timescales is more difficult. In principle, we can have many thermal modes with arbitrarily small thermal response times. However, the nature of the thermal susceptibility (\autoref{eq:ThermalSusceptibility}) is such that as $\tau_k$ increases the amplitude of the thermal modes decreases, in such a way that thermal modes for which $\Omega \tau_k\ll1$ are irrelevant. Other than this, the spatial profiles of faster thermal modes are often too complex, presenting small-scale spatial variations, in such a way that their overlap with the optical heating profile $R_k$ decreases very quickly as $\tau_k$ decreases. In practice, only a few thermal modes, with longer decay times, present a relevant overlap with the heating profile. We know from simulations that the response of those modes is much slower than the mechanical timescale $1/\tau_k\ll\Omega$. In this regime some simplifications can be made\footnote{ Here we have used the fact that in Eq.\ref{eq:thermal_resistance} $R^\theta_k$ is proportional to $\tau_k$ and the assumption $1/\tau_k\ll\Omega$.}:
\begin{equation}
  G_{\mathrm{pt},k}(\Omega) = \omega_0 \kappa_\text{abs}\frac{R^\theta_k\chi^{\theta}_k (\Omega)\Lambda^{\theta}_k}{\tau_k} \propto \frac{1}{\Omega^2\tau_k} + i \frac{1}{\Omega}.
  \label{eq:approximation}
\end{equation}
As such, we affirm that $\frac{\mathrm{Im}[G_{\mathrm{pt},k}(\Omega)]}{\mathrm{Re}[G_{\mathrm{pt},k}(\Omega)]} \approx \Omega\tau_k \gg 1$.

Expanding the imaginary part in \autoref{eq:delta_gamma_theta} in terms of the real and imaginary parts of $G_\mathrm{pt}(\omega)$ and $\Psi(\Omega,\Delta_0)$ we arrive at
\begin{align}
  \begin{split}
    \delta \Gamma^\theta =& -\frac{\hbar G_{x} |a_0|^2 }{ m_{\text{eff}} \Omega }\left( \text{Im}\left[G_\mathrm{pt}(\Omega)  \right] \text{Re}\left[\Psi(\Omega,\Delta_0)\right]\right. \\
    &\qquad \left.+ \text{Re}\left[G_\mathrm{pt}(\Omega)  \right] \text{Im}\left[\Psi(\Omega,\Delta_0)\right] \right) .
  \end{split}
\end{align}
The second term in the above expression is much smaller than the first because, as discussed previously, $\mathrm{Re}[G_{\mathrm{pt}}(\Omega)] \ll \mathrm{Im}[G_{\mathrm{pt}}(\Omega)]$ and $\text{Im}\left[ \Psi(\Omega,\Delta_0) \right] \ll \text{Re}\left[ \Psi(\Omega,\Delta_0) \right]$, and as such $\delta \Gamma^{\text{RP}}$. As such,
\begin{equation}
  \delta \Gamma^\theta \approx -\frac{\hbar G_{x} |a_0|^2 }{ m_{\text{eff}} \Omega } \text{Im}\left[G_\mathrm{pt}(\Omega)  \right] \text{Re}\left[\Psi(\Omega,\Delta_0)\right],
\end{equation}
indicating that for the photothermal backaction the dominant delay mechanism comes from the thermal dynamics.

Notice that in the regime $1/\tau_k\ll\Omega$ the size of the effect is nearly independent of the value of the thermal conductivity. This can be seen in \autoref{eq:approximation}, where the dominant term $i/\Omega$ is not dependent on $\tau_k$, which is the only parameter in our model dependent on thermal conductivity. In the regime considered here, we are assuming that within a single mechanical period heat has no time to flow. As such, the generated heat accumulates locally in such a way that the heating intensity is in quadrature with the temperature fluctuations.

\subsection{Model limitations}
As the temperature lowers, the overall number of phonons decreases, and with it phonon-phonon scattering interactions. As a consequence, the average phonon mean free path increases and with it silicon bulk conductivity. Indeed, the mean free path can become much larger than the device dimensions. In this case, the phonon scattering at the surface of the device becomes relevant. Usually, the scattering is diffuse because of surface roughness, characterizing the semi-ballistic regime, where the thermal behavior is yet diffusive, but the overall thermal conductivity depends on the device dimensions. At even lower temperatures, we can have specular scattering at the surface, when the typical phonon wavelength is much larger than the surface roughness. This defines the ballistic regime where heat transport is essentially wave-like.

Considering the temperature in which the experiments were realized, \SI{30}{K}, the estimated phonon mean free path\footnote{This is a rough estimate from the mean free path assuming the simplified expression $\kappa_\text{th} = \frac{1}{3} C_v v L$, where $C_v$ is the specific heat, $v$ is the average phonon speed in Silicon and $L$ is the mean free path.} is around \SI{60}{\micro\metre}, which is much larger than the typical size of our device \SI{220}{nm}, as such phonon scattering\footnote{Calculated from $\frac{v\hbar}{2\pi k_b T}$} is dominated by surface scattering. The typical phonon wavelength is around \SI{9}{nm}, which is of the same order of magnitude as the expected lateral surface roughness in silicon devices, which is around \SI{5}{nm}, indicating that we are still in the semi-ballistic regime, but ballistic effects are not completely negligible.

Using as the average width in the horizontal direction $\frac{170 + 500}{2}$\SI{}{nm} $ = $ \SI{335}{nm} and the thickness of \SI{220}{nm}, we arrive at a typical typical mean free path for surface scattering given by $\sqrt{335.220}$ \SI{}{nm} $ = $ \SI{270}{nm}. With this we can use the simplified expression $\kappa_\text{th} = \frac{1}{3} C_v v L$ to estimate an effective thermal conductivity for the device $\kappa_\text{th} = $ \SI{23}{W/(K.m)}.

Aware of possible limitations, the results of our model were used as a guide to design the structures at \SI{30}{K}. Ballistic effects related to differences in the surface properties between the inner and outer walls of the nanobeams could be one of the reasons behind the discrepancy between simulations and experiment.

\bibliographystyle{unsrt}
\bibliography{references}

\end{document}